\documentclass[twocolumn]{aastex631}

\usepackage{subfigure} 
\usepackage{amssymb}
\usepackage{amsmath}
\usepackage[para]{threeparttable}
\usepackage{hyperref}

\received{XXX}
\revised{YYY}
\accepted{ZZZ}

\shorttitle{Stellar Loci. IX. Estimation of Stellar Parameters from CSST-like Photometry}
\shortauthors{Lu et al.}

\begin{document}

\title{Stellar Loci. IX. Estimation of Stellar Parameters from CSST-like Photometry}

\author[0009-0009-0678-2537]{Xue Lu}
\affiliation{Institute for Frontiers in Astronomy and Astrophysics, Beijing Normal University,  Beijing 102206, China}
\affiliation{School of Physics and Astronomy, Beijing Normal University No.19, Xinjiekouwai St, Haidian District, Beijing, 100875, China}

\author[0000-0003-2471-2363]{Haibo Yuan}
\affiliation{Institute for Frontiers in Astronomy and Astrophysics, Beijing Normal University,  Beijing 102206, China}
\affiliation{School of Physics and Astronomy, Beijing Normal University No.19, Xinjiekouwai St, Haidian District, Beijing, 100875, China}

\author[0000-0001-8424-1079]{Kai Xiao}
\affiliation{Institute for Frontiers in Astronomy and Astrophysics, Beijing Normal University,  Beijing 102206, China}
\affiliation{School of Physics and Astronomy, Beijing Normal University No.19, Xinjiekouwai St, Haidian District, Beijing, 100875, China}
\affiliation{School of Astronomy and Space Science, University of Chinese Academy of Sciences, Beijing 100049, China}

\author[0000-0002-1259-0517]{Bowen Huang}
\affiliation{Institute for Frontiers in Astronomy and Astrophysics, Beijing Normal University,  Beijing 102206, China}
\affiliation{School of Physics and Astronomy, Beijing Normal University No.19, Xinjiekouwai St, Haidian District, Beijing, 100875, China}

\author[0000-0003-1863-1268]{Ruoyi Zhang}
\affiliation{Institute for Frontiers in Astronomy and Astrophysics, Beijing Normal University,  Beijing 102206, China}
\affiliation{School of Physics and Astronomy, Beijing Normal University No.19, Xinjiekouwai St, Haidian District, Beijing, 100875, China}

\author[0000-0002-9824-0461]{Lin Yang}
\affiliation{Department of Cyber Security, Beijing Electronic Science and Technology Institute, Beijing, 100070, China}

\author[0000-0003-4573-6233]{Timothy C. Beers}
\affiliation{Department of Physics and Astronomy, University of Notre Dame, Notre Dame, IN 46556, USA}
\affiliation{Joint Institute for Nuclear Astrophysics -- Center for the Evolution of the Elements (JINA-CEE), USA}

\author[0000-0003-3535-504X]{Shuai Xu}
\affiliation{Institute for Frontiers in Astronomy and Astrophysics, Beijing Normal University,  Beijing 102206, China}
\affiliation{School of Physics and Astronomy, Beijing Normal University No.19, Xinjiekouwai St, Haidian District, Beijing, 100875, China}

\correspondingauthor{Haibo Yuan}
\email{yuanhb@bnu.edu.cn}

\begin{abstract}
The China Space Station Telescope (CSST) will conduct a deep and wide imaging survey in the $NUV$-, $u$-, $g$-, $r$-, $i$-, $z$-, and $y$-bands.
In this work, using theoretical data synthesized from the BOSZ spectra of \cite{Bohlin2017}, along with observational data constructed from different sources, we present two methods for estimating stellar parameters from CSST-like photometry. One approach is to estimate metallicity [M/H] and surface gravity log\,\textit{g} simultaneously by using the metallicity- and log\,\textit{g}-dependent stellar loci.
Tests with theoretical data (without photometric errors) result in precisions of 0.088\,dex and 0.083\,dex for [M/H] and log\,\textit{g}, respectively.
With 0.01\,mag photometric errors, precision is degraded by about a factor of two, due to degeneracy in [M/H] and
log\,\textit{g}.
Tests with observational data, although with larger photometric errors, result in precisions of 0.10\,dex and 0.39\,dex for [Fe/H] and log\,\textit{g}, respectively, thanks to the strong 
correlation between stellar colors and log\,\textit{g} in real data. The other approach is the giant-dwarf loci method to obtain classifications and metallicity estimates.  With the same observational data, it achieves a better [Fe/H] precision of 0.084\,dex, due to the stronger constraints imposed on log\,\textit{g}. The method also performs well in distinguishing giants from dwarfs, particularly for red or metal-poor giants. 
This work demonstrates the clear potential of the CSST data, paving the way for stellar-parameter estimates for many billions of stars.

\end{abstract}


\section{Introduction} \label{sec:intro}

Stellar parameters are fundamental to studies of stellar physics, stellar populations, accreted structures, and the formation and evolution of the Milky Way (MW), as well as for exoplanets and interstellar dust. 
In the era of big data, there is a growing demand for larger samples of stars with more precise stellar parameters, especially for metallicity estimates.
At present, Stellar parameters are measured primarily using spectroscopic data. Thanks to the development of multi-object spectroscopy technique, a number of spectroscopic surveys e.g., SDSS/SEGUE (\citealt{SDSS2000,Yanny2009,Rockosi2022}), the Radial Velocity Experiment (RAVE; \citealt{RAVE2006}), the Large Sky Area Multi-Object Fiber Spectroscopic Telescope (LAMOST; \citealt{Deng2012,Liu2014}), the Apache Point Observatory Galactic Evolution Experiment (APOGEE; \citealt{Majewski2017}), the GALactic Archaeology with HERMES (GALAH; \citealt{GALAH2018}), the Gaia-ESO Public Spectroscopic Survey (\citealt{Gilmore_2022,Randich_2022}), and the Dark Energy Spectroscopic Instrument (DESI; \citealt{DESI2019,DESI2025}) have obtained such data for significant numbers of targets over the past two decades, yet they represent only a small fraction of the estimated total numbers of stars in the MW. 
Photometric surveys, in contrast, are much less expensive, can go deeper, and are significantly less biased; thus,
they provide a compelling alternative to deliver stellar parameters for huge numbers of stars.

The stellar-loci fitting method 
has been widely employed in photometric stellar-parameter determinations.
\citeauthor{yuan2015metal} (\citeyear{yuan2015metal}, Paper I) found that a 1\,dex decrease in metallicity of FGK main-sequence stars results in a decrease of about 0.20/0.02\,mag in the $u - g/g - r$ colors and an increase of about 0.02\,mag in the $r - i/i - z$ colors, respectively. They also found that the intrinsic widths of metallicity-dependent stellar loci are nil. As a result,  \citeauthor{yuan2015FGK} (\citeyear{yuan2015FGK}, Paper III) estimated photometric metallicities for half a million FGK stars from SDSS Stripe 82, with a typical precision of 0.1\,dex. \cite{Huang2019,Huang2022,Huang2023} applied a similar technique to the SkyMapper Southern Sky Survey (SMSS DR2; \citealt{SkyMapper_DR2}) and Stellar Abundances and Galactic Evolution Survey(SAGES DR1; \citealt{SAGES_DR1}), whose optimally designed narrow/medium-band $u$ and $v$ filters enabled photometric-metallicity estimates down to [Fe/H] $\sim -3.5$. In \citet{huang2025}, they present an updated stellar-parameter catalog for over fifty million stars from SMSS DR4 (\citealt{onken2024}).  

Although bluer and narrower bands are more sensitive to metallicity,  the availability of high-precision photometric data for redder and broader bands can also be used for precise photometric estimates of stellar metallicity. For instance, using the ultra broad-band Gaia Early Data Release 3 (\citealt{Brown2021}) photometry, \citeauthor{Xu_2022_metal} (\citeyear{Xu_2022_metal}, Paper V) obtained metallicity estimates with a typical precision of about 0.2\,dex for about 27 million stars. A 1\,dex change in [Fe/H] results in a 5\,mmag change in the $(BP-G)$ color for solar-type stars. Despite the very weak metallicity sensitivity, the exquisite data quality of $G_{BP}, G_{RP}, G$ allows very good precision. \citeauthor{Huang_bw2025} (\citeyear{Huang_bw2025}, Paper VIII) applied a similar method to synthetic $(BP–RP)_{XPSP}$ and $(BP–G)_{XPSP}$ colors derived from ``corrected" Gaia $BP/RP (XP)$ spectra, and obtained precise estimates of metallicity down to [Fe/H] $\sim -4$ for about 100 million stars in the MW. The typical metallicity precision is between 0.05 and 0.1\,dex for both dwarfs and giants at $\rm [Fe/H] = 0$, which is about three times better than previous work based on Gaia EDR3 colors in paper V. Note that \cite{xiao2024} searched for an optimal filter design for the estimation of stellar metallicity. That experiment showed that it is crucial to take into account uncertainty alongside the sensitivity when designing filters for measuring the stellar metallicity, and should lead to improved behavior in the near future.

At present, the GALEX $NUV$-band is the most sensitive filter to metallicity, as demonstrated by \citeauthor{Lu_2024} (\citeyear{Lu_2024}, Paper VII). It was found that a 1\,dex change in [Fe/H] results in a 1 mag change in the $(NUV-G_{BP})$ color for solar-type stars. Using GALEX and Gaia EDR3 data, Paper VII has estimated metallicities with a precision of 0.11\,dex for about 4.5 million dwarfs and 0.17\,dex for approximately 0.5 million giants. 
However, there are also challenges when using the $NUV$-band data for metallicity measurements, such as poor photometric precision, the effects of stellar activity, and high sensitivity to extinction. A further challenge lies in the strong dependence of the $NUV$-band on surface gravity (or absolute magnitude), which has to be taken into account in modeling.

The China Space Station Telescope (CSST; \citealt{zhan_2011,zhan_2021,CSST2025}), with a 2 m aperture and a field of view of 1.1 degrees, will conduct deep, high-resolution and large-area multi-band imaging and slitless spectroscopic surveys covering the wavelength range of 255--1000 nm. 
For the multi-band imaging survey, it will image roughly 17,500 square degrees of the sky in the $NUV$-, $u$-, $g$-, $r$-, $i$-, $z$-, and $y$-bands. The 5$\sigma$ limiting magnitude in the $NUV$-band can reach 25 (AB magnitudes) for point sources. Such a massive amount of data will provide a great opportunity to estimate precise stellar parameters for an enormous number of stars. 
However, due to the lack of distance information, the technique developed in \cite{Lu_2024} cannot be applied directly to most CSST stars. Thus, in this work, we aim to explore alternative stellar-parameter estimation techniques for use with the upcoming CSST data.

Previously, \cite{shi_2024} studied the performance of CSST broadband photometry to estimate stellar atmospheric parameters. They first used the $NUV-u$ vs. $g-i$ color--color diagram to classify dwarfs and giants, and then employed the metallicity-dependent stellar locus of colors $u-g$ and $g-i$ to estimate metallicity.  However, both the $NUV$- and $u$-bands are highly sensitive not only to metallicity but also to surface gravity (log\,\textit{g}).
Taking this into account, we propose an optimal
metallicity- and log\,\textit{g-dependent}-dependent stellar-loci method to simultaneously estimate metallicity and log\,\textit{g} based on CSST-like filters in this work. However, we find that the degeneracy in metallicity and log\,\textit{g} makes it difficult to estimate these parameters accurately, particularly in cases of large photometric errors. Therefore, we also explore the giant-dwarf loci method to classify giants and dwarfs, and then estimate their photometric metallicities.

This paper is organized as follows. Sections \ref{sec:data} and \ref{sec:method} describe the data and methods used in this work. Sections \ref{sec:model_data} and \ref{sec:observ_data} present the tests of the metallicity- and log\,\textit{g}-dependent stellar-loci method with theoretical (simulated) and observational (real) data, respectively. Section\,\ref{sec:select_giant} presents tests of the giant-dwarf loci method with observational data. We summarize in Section \ref{sec:summary}.

\section{Data} \label{sec:data}
In this work, we employ two sets of data to mimic the CSST photometry: a theoretical set and an observational set. The former relies on synthetic colors from theoretical spectra, as introduced in section\,\ref{subsec:m_data}. In the latter case, 
we use ``real" data from different sources, as described in sections\,\ref{subsec:galex} - \ref{subsec:lamost}.

\subsection{Theoretical Data} \label{subsec:m_data}
The theoretical data we use consist of synthetic CSST colors for various atmospheric parameters, calculated from the BOSZ stellar spectra (\citealt{Bohlin2017}) convolved with the CSST transmission curves (Figure\,7 of \citealt{zhan_2021}).
The BOSZ spectra are based on the ATLAS9-APOGEE (\citealt{Sz2012}) and MARCS (\citealt{Gustafsson2008}) model atmospheres and cover a wide range of effective temperature, surface gravity, metallicity, carbon abundance, and alpha-element abundance. Two versions of these spectra are available, the original 2017 version and a newer 2024 version with different resolution. 
Here we use the 2017 version spectra; the parameter ranges can be seen in the tables and Figure\,1 in \cite{Bohlin2017}.
The parameter [M/H] ranges from $-$2.5 to $+$0.5 at a step size of 0.25\,dex, while log\,\textit{g} ranges from 0.0 to 5.0 with a step size of 0.5\,dex.
With the assumption that $\rm [\alpha/M]=0$, $\rm [C/M]=0$, and $-0.4<g-i<1.5$, our theoretical data sample includes 2312 spectra.

The $\rm [\alpha/M]$ and [C/M] are not considered because the CSST broadband filters used in this work are not sensitive enough to constrain these elemental abundances. Quantitative determinations of $\rm [\alpha/M]$ and [C/M] would require either dedicated narrow-band filters or the slitless spectroscopy expected from CSST. However, for carbon-enhanced very metal-poor stars, it is still possible to identify and estimate [C/M] using the CSST $NUV$ and $u$ bands, since the $NUV$-band is free from molecular carbon absorption features that strongly affect the $u$-band. Moreover, as the $NUV$-band is also unaffected by molecular nitrogen absorption features, which are primarily concentrated in the $u$-band, it may also be used to search for nitrogen-enhanced metal-poor stars. (Yu L. et al., in prep.).

\subsection{GALEX GR5 Medium-depth Imaging Survey} \label{subsec:galex}
The Galaxy Evolution Explorer (GALEX), as the first mission to conduct a space-based sky survey in both the near and far ultraviolet, includes an all-sky imaging survey, and medium and deep imaging surveys covering specific areas, as well as spectroscopic surveys utilizing grism technology. It simultaneously performs sky surveys in two bands through the use of a dichroic beam splitter: far-UV ($FUV$; $1344-1786 \rm \AA$) and near-UV ($NUV$; $1771-2831 \rm \AA$).
The GALEX GR5 Medium-depth Imaging Survey (MIS) database (\citealt{Bianchi2011}) contains 12.6 million sources, covers 1579 square degrees, and has a depth in $FUV/NUV$ of 22.6/22.7 (AB magnitudes). 
The stars roughly 21st magnitude stars in the $NUV$-band have a median photometric error of 0.05\,mag.

\vskip 1cm
\subsection{Synthetic SDSS Magnitudes from ``Corrected'' Gaia XP Spectra} \label{subsec:gaia}

Gaia Data Release 3 (DR3; \citealt{gaia_collaboration2023}) provides astrometric measurements for nearly two billion stars. Additionally, Gaia DR3 introduces, for the first time, low-resolution spectra from the Blue and Red Photometer (BP and RP) observations. These spectra cover the optical to near-infrared wavelength range of 330 to 1050 nm. They are available for approximately 220 million sources, primarily those brighter than $G$ = 17.65.  
The ``corrected'' Gaia XP Spectra we employ is from \cite{Huang_bw2024}. They have analyzed and corrected the color-, magnitude-, and reddening-dependent systematic errors in the Gaia DR3 XP spectra by combining CALSPEC (\citealt{calspec2022}) and NGSL (\citealt{ngsl2007}), as well as the spectroscopic data from LAMOST Data Release 8 (DR8; \citealt{Luo2015}). Convolving the ``corrected'' Gaia XP spectra with the SDSS transmission curves (\citealt{Fukugita1996}), we obtain synthetic $u,g,r,i,z$ magnitudes.

\subsection{LAMOST Data Release 8} \label{subsec:lamost}
The Large Sky Area Multi-Object Fiber Spectroscopic Telescope (LAMOST; \citealt{Luo2015}) is a 4-meter quasi-meridian reflecting Schmidt telescope equipped with 4000 fibers. In this study, we used the LAMOST DR8\footnote{\href{https://www.lamost.org/dr8/v2.0/}{https://www.lamost.org/dr8/v2.0/}} low-resolution catalog, which contains 10,809,336 spectra and stellar parameters for over six million stars. The parameters from LAMOST were derived using the LASP pipeline (\citealt{Wuyue2014}), based on the ELODIE stellar spectral library. The precision of these parameters has been validated through repeated observations and extensive comparisons with external high-resolution datasets (see \citealt{Luo2015} for details).
 
\subsection{Construction of the Observational Data} 
\label{subsec:observ-sample}
We combine the GALEX GR5 MIS, synthetic SDSS magnitudes from the ``corrected'' Gaia XP Spectra, and the LAMOST DR8 dataset to construct our observational data. In total, 137,990 sources are matched with a $1^{\prime\prime}$ cross-matching radius.
We further apply the following criteria to select 
high-quality sample stars: 
\begin{enumerate}

\item[1)] $0.2<(g-i)_0<1.5,\ (g-i)_0=(g-i-R_{g-i}\times E(B-V))$;

\item[2)] $\rm NUV>15$ to avoid saturation (\citealt{Morrissey_2007(galex_saturation)}) and $\rm error_{NUV}<0.1$ to ensure the $NUV$-band data quality;

\item[3)] LAMOST spectral signal-to-noise ratio in the $g$-band $\geq 20$;

\item[4)] $\rm phot\_bp\_rp\_excess\_factor < 0.105 \times (BP-RP)+1.15$ to ensure the data quality of the Gaia XP spectra; 

\item[5)] $E(B-V)\leq0.025$ to minimize uncertainties due to reddening corrections. The $\rm E(B-V)$ used here is from the \citeauthor{SFD98} (\citeyear{SFD98}, hereafter SFD98) dust reddening map, which is one of the best choices for reddening correction of stars at middle and high Galactic latitudes (\citealt{Sun_2022}).

\end{enumerate}

\begin{figure}[htbp]
\centering
\includegraphics[width=1.0\linewidth]{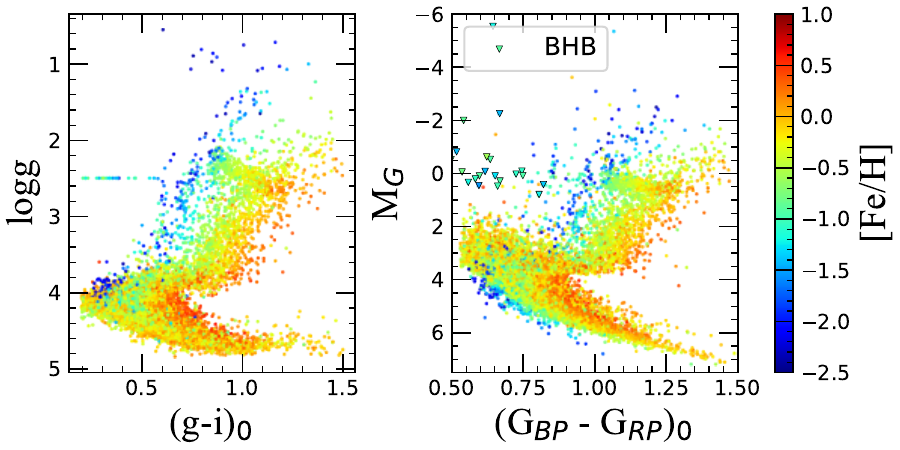}
\caption{H-R diagram for the observational data, shown as ($g-i$)$_0$ -- log\,\textit{g} and (G$_{BP}$-G$_{RP}$)$_0$ -- M$_G$, color-coded by [Fe/H] shown in the color bar on the right. The BHB stars selected from the right panel are assigned a log\,\textit{g} value of 2.5. }
\label{fig:sample}
\end{figure}

Finally, 13,267 sources remain as the observational sample, as shown in Figure \ref{fig:sample}, with a median photometric error of 0.07, 0.015, 0.008, 0.009, 0.012\,mag in the $u$-, $g$-, $r$-, $i$-, $z$-band. The Note that there are some issues with the log\,\textit{g} parameter for Blue horizontal-branch (BHB) stars in the LAMOST data. As a result, we select BHB stars from the color-magnitude diagram and assign them a log\,\textit{g} value of 2.5.

\section{Methods} \label{sec:method}

\subsection{Reddening Corrections} \label{subsec:redden}
For all observational data, careful reddening corrections are performed using the $T_{ eff}$- and $E(B-V)$-dependent reddening coefficients from the python package of \cite{zhang_2023}. All colors referred to hereafter are the intrinsic (de-reddened) colors, unless otherwise noted.
In Paper VII, we found differences in the reddening coefficients between dwarfs and giants. Such differences are ignored here, considering that  we only selected sources with low extinction values of $E(B-V)<0.025$.

\subsection{[Fe/H]-dependent Stellar Loci} \label{subsec:feh_depen}

Following \citet{yuan2015metal}, the [Fe/H]-dependent stellar-loci method conducts a two-dimensional polynomial fitting for $NUV-g$, $u-g$, $g-r$, $i-z$ and $z-y$ (no $y$-band in the observational data sample), as a function of $g-i$ color and metallicity [Fe/H], and employs a minimum $\chi^2$ technique to derive metallicity for dwarfs and giants, respectively. 
For dwarfs, a fourth-order polynomial with 15 free parameters is adopted for the $NUV-g$ color and a third-order polynomial with 10 free parameters is adopted for the colors $u-g$, $g-r$, and $i-z$. For giants, a fourth-order polynomial with 15 free parameters is adopted for the $u-g$ color and a third-order polynomial with 10 free parameters is adopted for the colors $NUV-g$, $g-r$, and $i-z$. A $3\sigma$ clipping is performed during all fitting processes; for $NUV-g$ we employ a $1.5\sigma$ clipping to remove the likely active stars.

The $\chi^2$ is defined as:
\begin{eqnarray}\label{chi2_1}
&&\chi^2(\rm [Fe/H]) = \nonumber \\ 
&&\sum \limits _{i=1} ^{4} \frac{[\rm C^{obs}_{i}- R_{c}^i\cdot E(B-V) - C^{int}_{i}(g-i, [Fe/H])]^2}{(\sigma_{c}^i)^2 +\sigma_{\rm g-i}^2},\nonumber \\ 
\end{eqnarray}
where $\rm C^{obs}_{i}$ is the observed color, and i = 1, 2, 3, 4 for colors $NUV-g$, $u-g$, $g-r$, and $i-z$, respectively. $R_{c}^i$ is the reddening coefficient for $\rm C^{obs}_{i}$, $E(B-V)$ is the extinction value of SFD98, $\rm C^{int}_{i}(g-i, [\rm Fe/H])$ represents the predicted intrinsic color based on the stellar loci dependence on [Fe/H], and $\sigma_{c}^i$ and $\sigma_{\rm g-i}$ represent the uncertainties in $\rm C^{obs}_{i}$ and $g-i$, respectively. 

\subsection{[Fe/H]- and log\,\textit{g}-dependent Stellar Loci} \label{subsec:feh_logg_depen}

To take into account both the effects of [Fe/H] (or [M/H]) and log\,\textit{g}, we conduct a three-dimensional polynomial fitting for stellar loci. 
For colors $NUV-g$, $u-g$, $g-r$, $i-z$ and $z-y$ of the theoretical data, a fifth-order polynomial with 38 free parameters is adopted. For the colors $NUV-g$, $u-g$, $g-r$, and $i-z$ from the observational data, a fourth-order polynomial with 31 free parameters is adopted.

The minimum $\chi^2$ technique is then used to simultaneously derive estimates of the metallicity and log\,\textit{g}.
The $\chi^2$ is defined as:
\begin{eqnarray}\label{chi2_2}
&&\chi^2(\rm [Fe/H], log\,\textit{g}) = \nonumber \\ 
&&\sum \limits _{i=1} ^{5(or\ 4)} \frac{[\rm C^{obs}_{i}- R_{c}^i\cdot E(B-V) - C^{int}_{i}(g-i, [Fe/H], log\,\textit{g})]^2}{(\sigma_{c}^i)^2 +\sigma_{\rm g-i}^2},\nonumber \\ 
\end{eqnarray}
where $\rm C^{int}_{i}(g-i, [Fe/H], log\,\textit{g})$ represents the predicted intrinsic color based on the [Fe/H]- and log\,\textit{g}-dependent stellar loci; the others are similar to Equation\,\ref{chi2_1}.

\section{Test of the [Fe/H]- and log\,\textit{g}-dependent stellar-loci method with synthetic data} \label{sec:model_data}

\begin{figure*}[htbp]
\centering
\includegraphics[width=16cm]{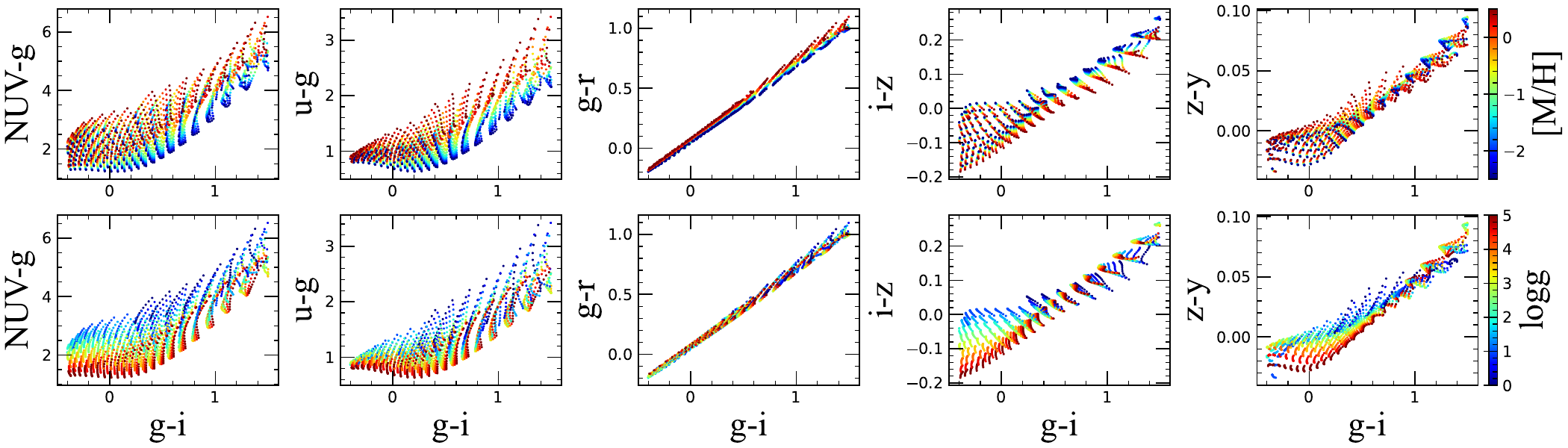}
\caption{{\small Stellar loci of theoretical CSST colors, color-coded by [M/H] and log\,\textit{g} shown in the color bars to the right of the upper and lower rows of panels, respectively.
}}
\label{Fig:model-ccd}
\end{figure*}

The stellar loci of synthetic data are shown in Figure\,\ref{Fig:model-ccd}. The dependence of not only [M/H] but also log\,\textit{g} can be seen in these color-color diagrams. The $NUV$- and $u$-band are sensitive to both [M/H] and 
log\,\textit{g}, as expected. 

To use the [Fe/H]- and log\,\textit{g}-dependent stellar-loci method,
a fifth-order polynomial with 38 free parameters is adopted for colors the $NUV-g$, $u-g$, $g-r$, $i-z$, and $z-y$. The relation is as follows:
\begin{eqnarray}\label{model-rela}
&Color =  p0\cdot X^5 + p1\cdot X^4\cdot Y + p2\cdot X^4\cdot Z \nonumber \\
&{}+ p3\cdot X^3\cdot Y^2 + p4\cdot X^3\cdot Z^2 + p5\cdot X^2\cdot Y^3 \nonumber \\
&{}+ p6\cdot X^2\cdot Z^3 + p7\cdot X^4 + p8\cdot Y^4 + p9\cdot Z^4 \nonumber \\
&{}+ p10\cdot X^3\cdot Y + p11\cdot Y^3\cdot X + p12\cdot X^3\cdot Z \nonumber \\
&{}+ p13\cdot Z^3\cdot X + p14\cdot Z^3\cdot Y + p15\cdot Y^3\cdot Z \nonumber \\
&{}+ p16\cdot X^2\cdot Y^2 + p17\cdot X^2\cdot Z^2 + p18\cdot Z^2\cdot Y^2 \nonumber \\
&{}+ p19\cdot X^3 + p20\cdot Y^3 + p21\cdot Z^3 + p22\cdot X^2\cdot Y \nonumber \\
&{}+ p23\cdot Y^2\cdot X + p24\cdot X^2\cdot Z + p25\cdot Z^2\cdot X \nonumber \\
&{}+ p26\cdot Z^2\cdot Y + p27\cdot Y^2\cdot Z + p28\cdot X^2 \nonumber \\
&{}+ p29\cdot Y^2 + p30\cdot Z^2 + p31\cdot X\cdot Y + p32\cdot Y\cdot Z \nonumber \\
&{}+ p33\cdot X\cdot Z + p34\cdot X + p35\cdot Y + p36\cdot Z + p37, \nonumber \\
\end{eqnarray}
where $X, Y, Z$ represents the color $g-i$, [M/H], and log\,\textit{g}, respectively. 

\begin{figure*}[htbp]
\centering
\subfigure{
\includegraphics[width=0.49\textwidth]{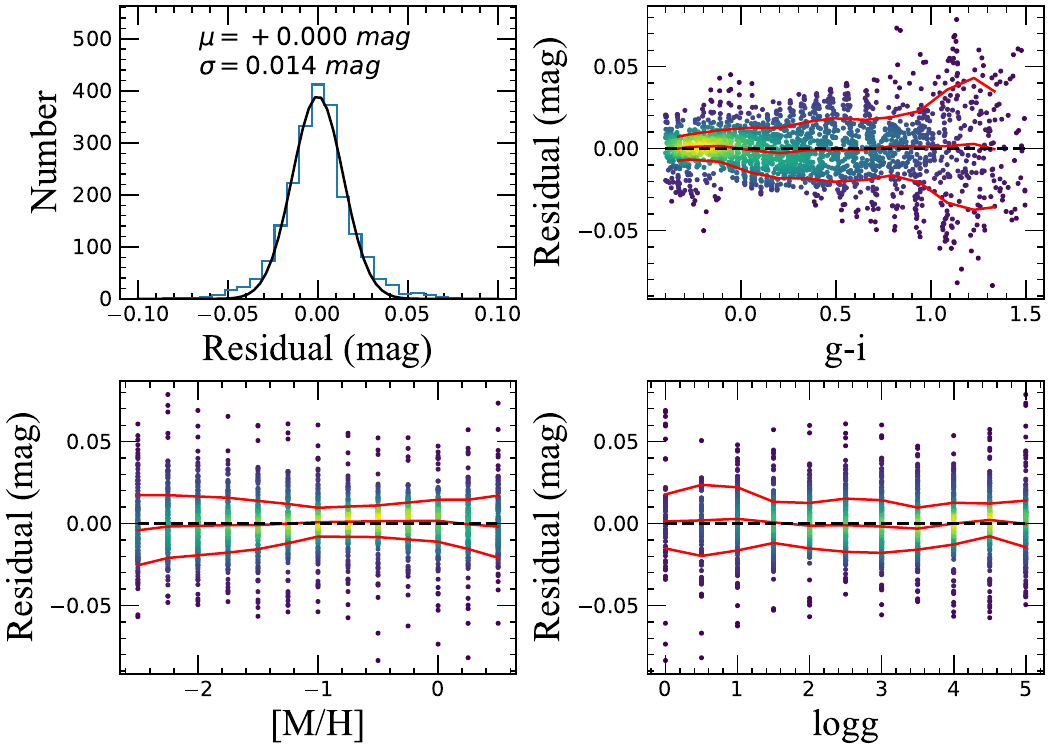}}
\subfigure{
\includegraphics[width=0.49\textwidth]{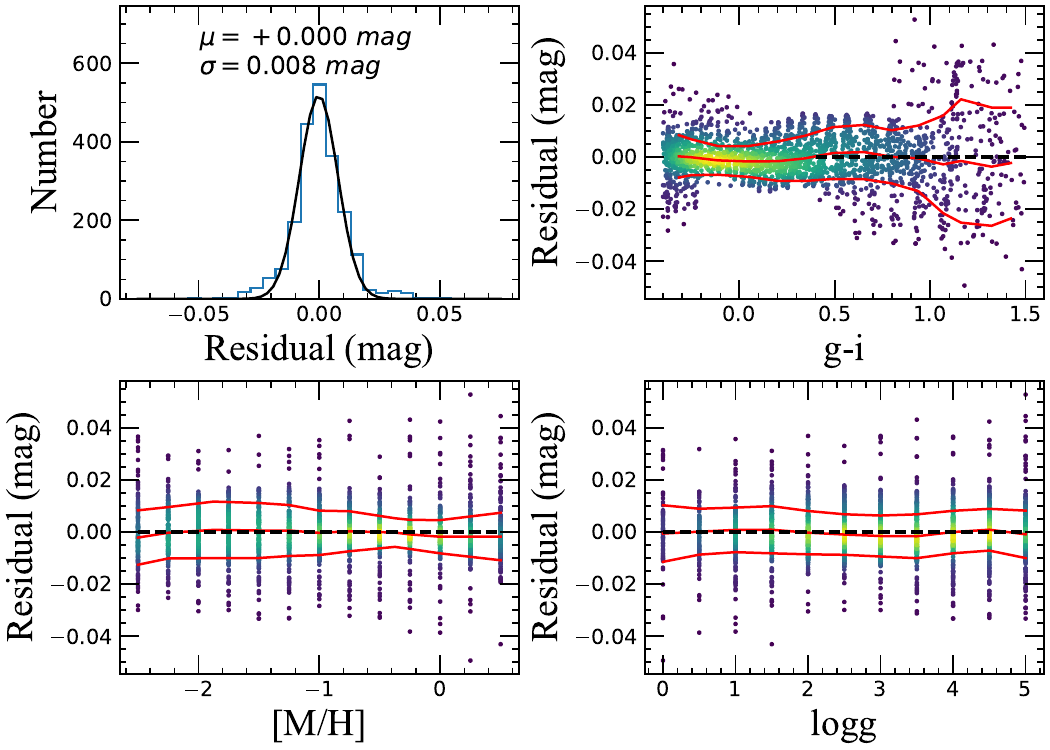}}
\caption{Fitting residuals of the [Fe/H]- and log\,\textit{g}-dependent stellar loci. The left four panels are for the $NUV-g$ color, which contain the histogram distribution of fitting residuals, with the Gaussian fitting profile over-plotted in black, and fitting residuals as a function of $g-i$, [M/H] and log\,\textit{g}, with the median values and standard deviations over-plotted in red.
The right four panels are similar to the left ones but for the $u-g$ color. The points are color-coded by their number density. }
\label{fig:simu_fit}
\end{figure*}

The fitting residuals for the $NUV-g$ and $u-g$ colors are shown in Figure\,\ref{fig:simu_fit}. 
The mean values of the fitting residuals are very close to 0.0 mag for all colors. The standard deviations are 0.014, 0.008, 0.002, 0.002, 0.001\,mag for the colors $NUV-g$, $u-g$, $g-r$, $i-z$, and $z-y$, respectively. 
The residuals exhibit no discernible trends with $g-i$, [M/H], and log\,\textit{g}.

\begin{figure*}[htbp] \centering
\includegraphics[width=16cm]{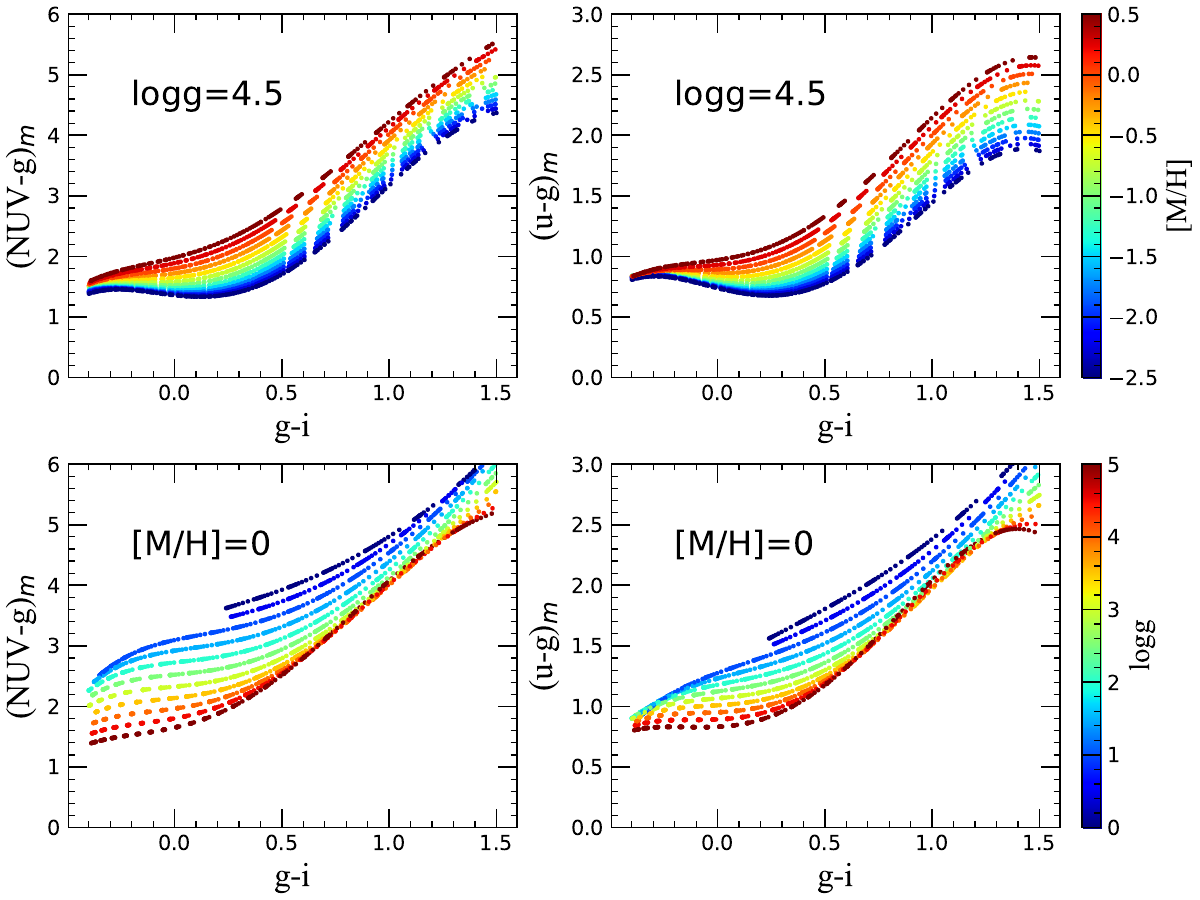}
\caption{
Top panels: The dependence of the $(NUV-g)_m$ (left) and $(u-g)_m$ (right) colors on [M/H], color-coded by [M/H] in the color bars to the right.
Bottom panels: The dependence 
of the $(NUV-g)_m$ (left) and $(u-g)_m$ (right) colors on log\,\textit{g}, color-coded by log\,\textit{g} in the color bars to the right. 
}
\label{fig:purity-ccd}
\end{figure*}

To examine the sensitivity to [M/H] and log\,\textit{g} independently, based on our models, the top panels of Figure\,\ref{fig:purity-ccd} plot the dependence of $(NUV-g)_m$ and $(u-g)_m$ colors on [M/H] for stars with $\rm log\,\textit{g} = 4.5$, and the bottom panels plot the dependence 
of the $(NUV-g)_m$ and $(u-g)_m$ colors on log\,\textit{g} for stars of $\rm [M/H] = 0.0$. Here, $(NUV-g)_m$ and $(u-g)_m$ are the colors predicted by our models.
From inspection, a 1\,dex change in metallicity corresponds to an approximately 0.5\,mag change in $NUV-g$ color and 0.25\,mag change in $u-g$ for solar-type stars. A 1\,dex change in log\,\textit{g} corresponds to an approximately 
0.2\,mag change in the $NUV-g$ color and a 0.1\,mag change in $u-g$ color for solar-type stars at $g-i=0.5$.

\begin{figure*}[htbp]
\centering
\includegraphics[width=13cm]{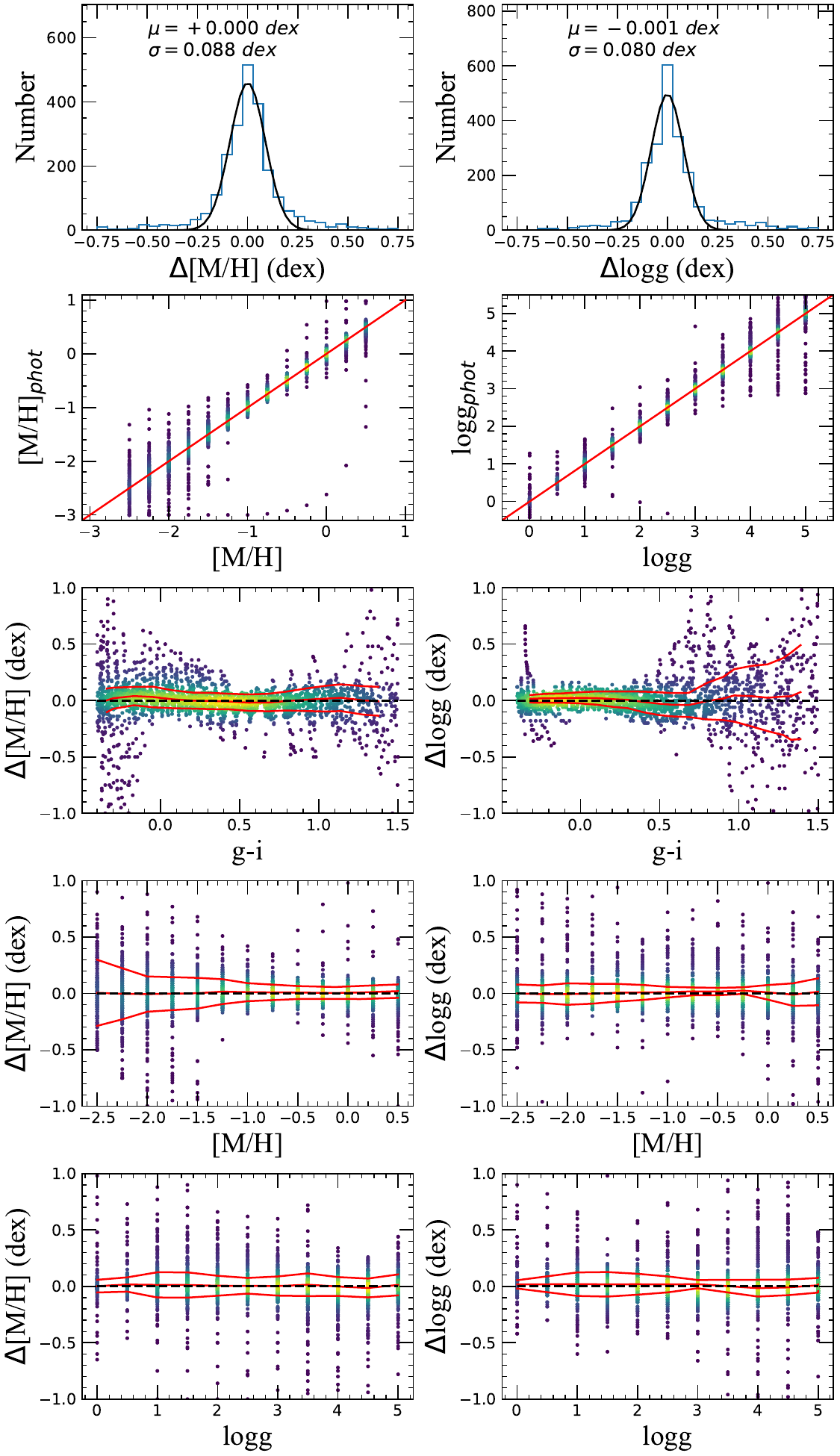}
\caption{
Left panels: The result for photometric-metallicity estimates [M/H]. From top to bottom, the panels show histogram distributions of the residuals $\Delta [\rm M/H]$, with the Gaussian fitting profile over-plotted in black,
comparison of the photometric-metallicity estimates with input values, $\Delta [\rm M/H]$ as a function of $g-i$ color, [M/H], and log\,\textit{g}. Right panels: Similar to the left ones but for log\,\textit{g}. The points are color-coded by their number density.
}
\label{fig:simu_result}
\end{figure*}

Based on the five relations above, we estimate the photometric metallicities and values of log\,\textit{g} using the minimum $\chi^2$ technique mentioned in Section \ref{subsec:feh_logg_depen}. We utilize a brute-force algorithm to determine the optimal [M/H] and log\,\textit{g} values for each source. For a given source, the value of [Fe/H] varies from $-$3.0 to +1.0 with steps of 0.02\,dex, and the value of log\,\textit{g} varies from $-$0.5 to $+$5.5 with steps of 0.02\,dex, resulting in values of 201$\times$301 $\chi^2$. The minimum $\chi^2$ among these values corresponds to the optimal [Fe/H] and log\,\textit{g} values. The results are shown in Figure\,\ref{fig:simu_result}. We have achieved a typical precision of 0.088\,dex for [M/H] and 0.083\,dex for log\,\textit{g}, respectively. 
Both residuals are flat with $g-i$ color, [M/H] and log\,\textit{g}. 
The $\rm [M/H]_{phot}$ errors are larger for more metal-poor stars due to 
their lower sensitivity to [M/H], as expected. The log\,\textit{g} errors are larger for redder stars due to their lower sensitivity on log\,\textit{g}.

\begin{figure*}[htbp]
\centering
\subfigure{
\includegraphics[width=4.25cm]{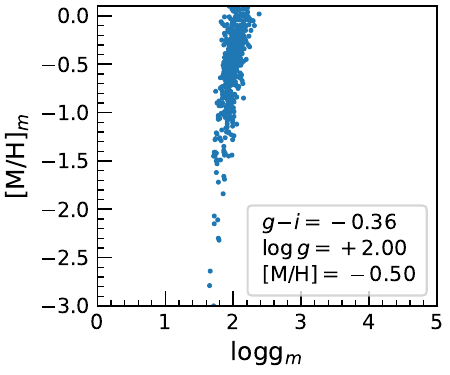}}
\subfigure{
\includegraphics[width=4.25cm]{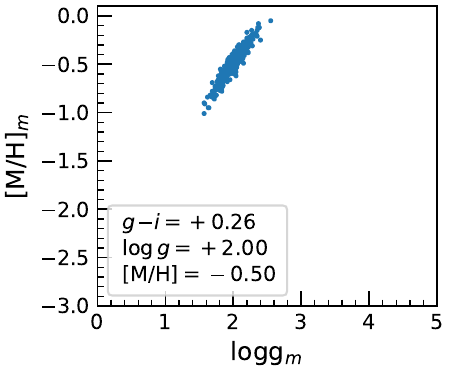}}
\subfigure{
\includegraphics[width=4.25cm]{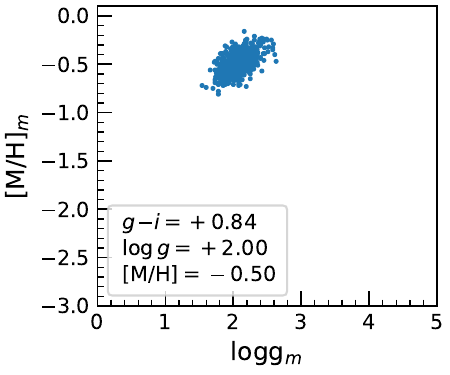}}
\subfigure{
\includegraphics[width=4.25cm]{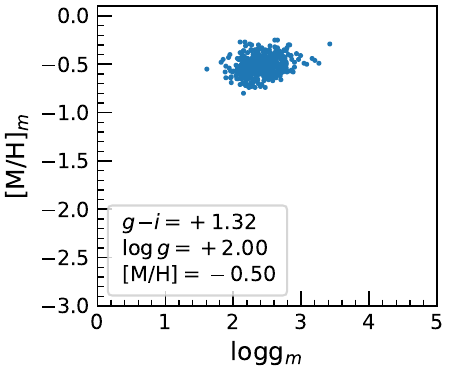}}
\subfigure{
\includegraphics[width=4.25cm]{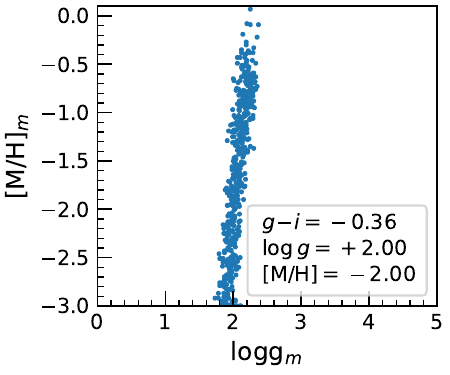}}
\subfigure{
\includegraphics[width=4.25cm]{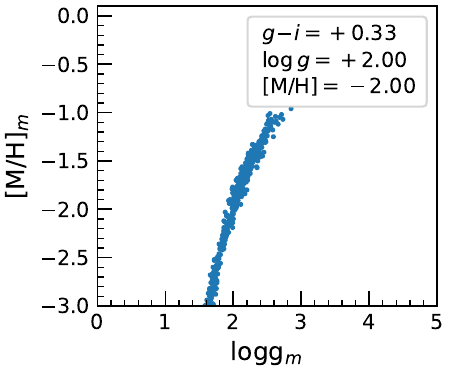}}
\subfigure{
\includegraphics[width=4.25cm]{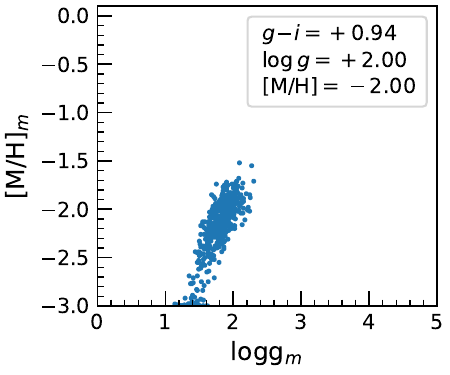}}
\subfigure{
\includegraphics[width=4.25cm]{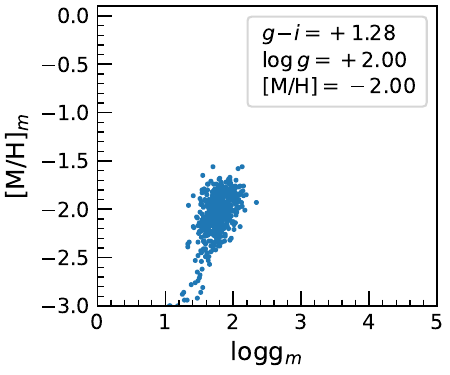}}
\subfigure{
\includegraphics[width=4.25cm]{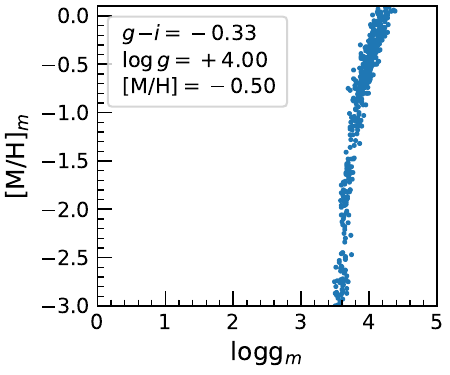}}
\subfigure{
\includegraphics[width=4.25cm]{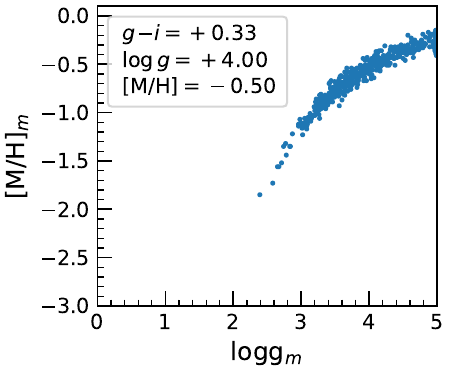}}
\subfigure{
\includegraphics[width=4.25cm]{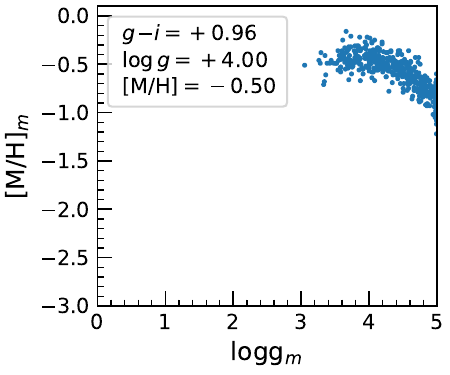}}
\subfigure{
\includegraphics[width=4.25cm]{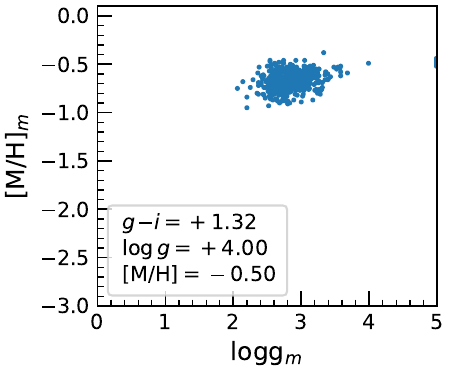}}
\subfigure{
\includegraphics[width=4.25cm]{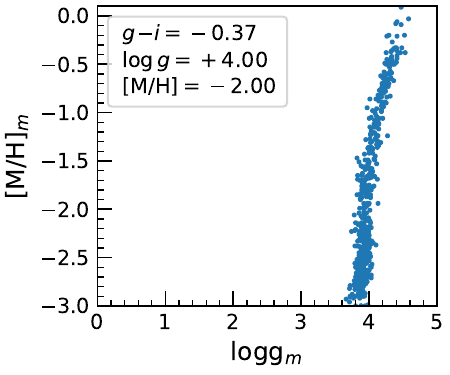}}
\subfigure{
\includegraphics[width=4.25cm]{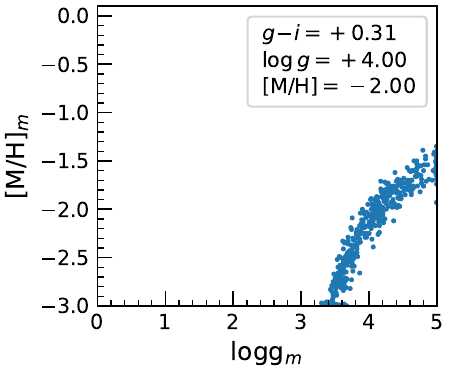}}
\subfigure{
\includegraphics[width=4.25cm]{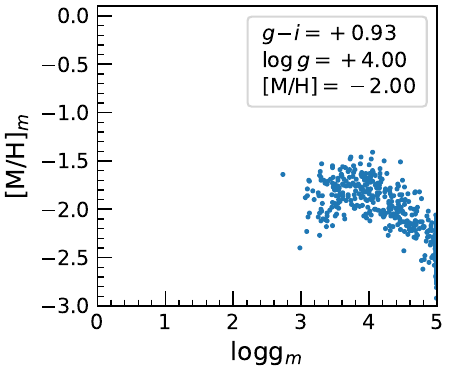}}
\subfigure{
\includegraphics[width=4.25cm]{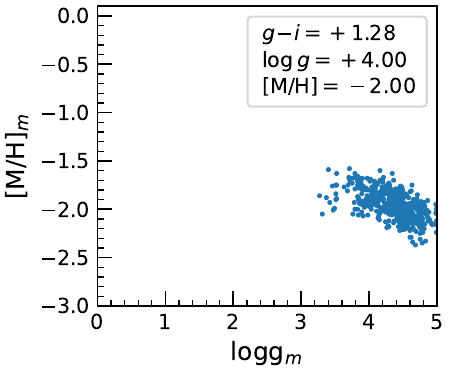}}

\caption{Monte Carlo simulation results for $\rm [M/H]$ and $\rm log\,\textit{g}$ estimates for stars with assumed 0.01 mag photometric errors and different parameters, as labeled in each panel. Panels from left to right: $g-i$ from bluer to redder. The top two rows show giants with $\rm log\,\textit{g}=2$, The bottom two rows show dwarfs with $\rm log\,\textit{g}=4$. The first and third rows have $\rm [M/H]=-0.5$, the second and fourth rows have $\rm [M/H]=-2.0$. }
\label{fig:simu_monte}
\end{figure*}

\begin{figure*}[htbp]
\centering
\includegraphics[width=14cm]{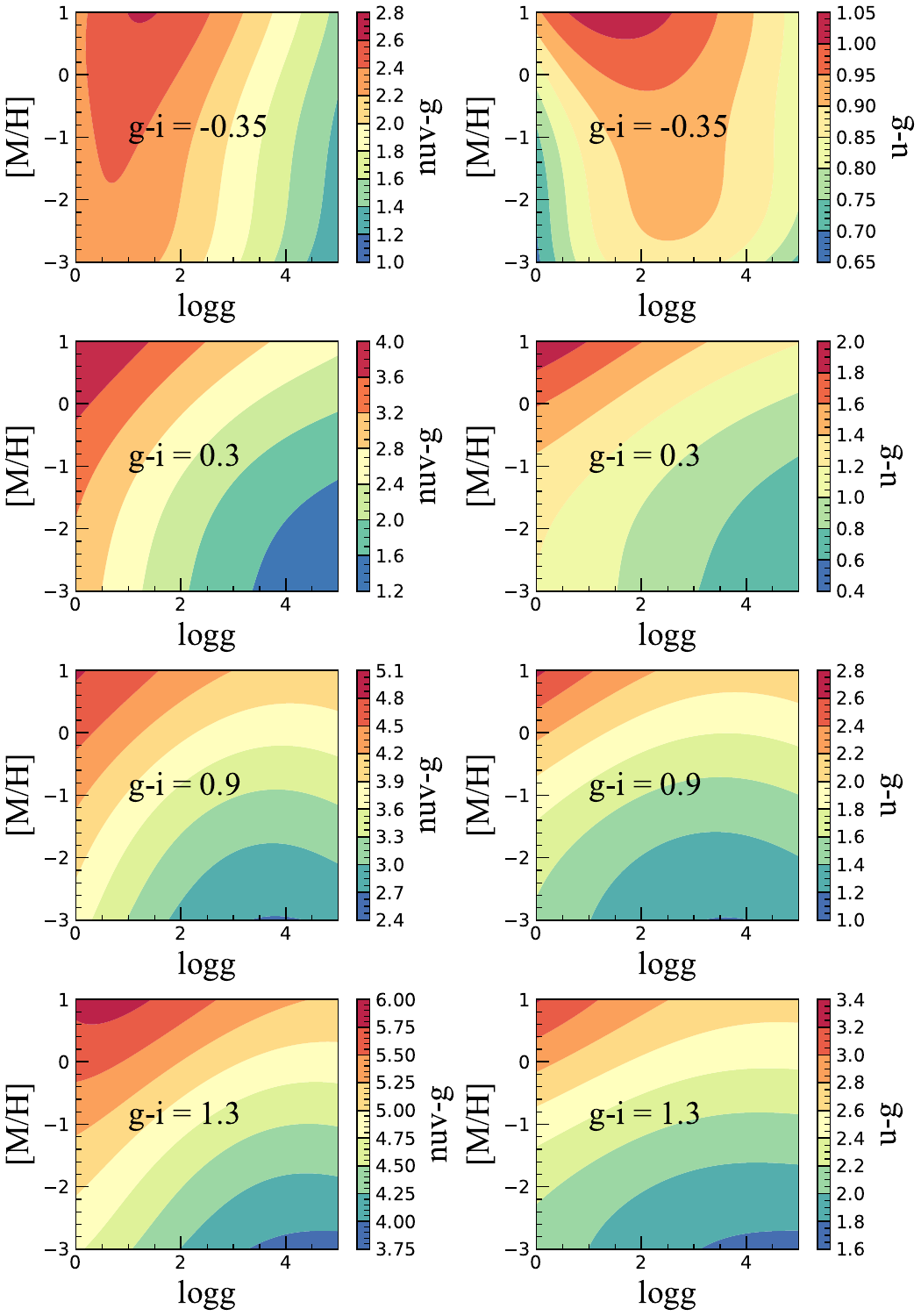}
\caption{$NUV-g$ (left panels) and $u-g$ (right panels) colors as a function of [M/H] and log\,\textit{g} for different $g-i$ colors as labeled on each panel. The $g-i$ colors closely match Figure\,\ref{fig:simu_monte}.}
\label{fig:coutour}
\end{figure*}

To investigate the effect of photometric errors on the derived parameters, we add a 0.01\,mag Gaussian photometric error to the theoretical data. The standard deviations of the fitting residuals for the colors $NUV-g$, $u-g$, $g-r$, $i-z$ and $z-y$ increase to 0.031\,mag, 0.022\,mag, 0.013\,mag, 0.015\,mag and 0.014\,mag, respectively.
For the photometric-parameter estimates, the typical [M/H] and log\,\textit{g} errors increase to 0.183\,dex and 0.176\,dex, respectively. 
The relatively larger errors are mainly caused by the degeneracy between metallicity and log\,\textit{g}.
To demonstrate this effect, we performed Monte Carlo simulations for 16 stars in our sample with different stellar parameters. For each star, Figure\,\ref{fig:simu_monte} plots the distribution of the estimated stellar parameters for 500 simulations with photometric errors of 0.01\,mag. Strong degeneracies are found in many cases, causing larger uncertainties in [M/H] and log\,\textit{g}, and even incorrect classifications between dwarfs and giants in some cases (such as the two middle columns of Figure\,\ref{fig:simu_monte}). We present Figure\,\ref{fig:coutour} as a complementary view, showing the contours of $NUV-g$ and $u-g$ colors as functions of [M/H] and log\,\textit{g} for the $g-i$ colors corresponding to the four columns in Figure\,\ref{fig:simu_monte}. The $NUV-g$ and $u-g$ colors are derived from our models, and a steeper color gradient indicates higher sensitivity to the corresponding parameter.
The correlation patterns in the simulation (Figure\,\ref{fig:simu_monte}) are consistent with those seen in the contour plots. For the first column ($g-i\sim -0.3$), the colors are weakly sensitive to metallicity but strongly dependent on log\,\textit{g} (see also Figure\,\ref{fig:purity-ccd}); thus, the uncertainties are dominated by [M/H], while log\,\textit{g} is well constrained. For the second column ($g-i\sim 0.3$), both [M/H] and log\,\textit{g} are sensitive, but their strong degeneracy leads to considerable uncertainties in both. For the third column ($g-i\sim 0.9$), both [M/H] and log\,\textit{g} are well determined for giants, whereas for dwarfs, the low sensitivity to log\,\textit{g} results in larger uncertainties, and an anti-correlation between [M/H] and log\,\textit{g} is observed. For the last column ($g-i\sim 1.3$), both [M/H] and log\,\textit{g} are well-constrained for giants, but dwarfs exhibit larger uncertainties in log\,\textit{g} for the same reason.

The degeneracy in metallicity and log\,\textit{g} makes it difficult to estimate both parameters accurately. 
This suggests two ways to improve -- one is the improvement of the photometric precision, the other is to break the degeneracy using priors. 

\section{Test of the [Fe/H]- and log\,\textit{g}-dependent stellar-loci method with observational data} \label{sec:observ_data}

The strong correlation between log\,\textit{g}, and colors in the observational data provides a strong prior that would help break up the metallicity-log\,\textit{g} degeneracy. Thus, the observational data is used for subsequent testing.

Here a fourth-order polynomial with 31 free parameters is adopted for the colors $NUV-g$, $u-g$, $g-r$, and $i-z$. The relation is:
\begin{eqnarray}\label{eq1}
&Color = \nonumber \\
&p0\cdot X^4 + p1\cdot Y^4 + p2\cdot Z^4 + p3\cdot X^3\cdot Y \nonumber \\
&{}+ p4\cdot Y^3\cdot X + p5\cdot X^3\cdot Z + p6\cdot Z^3\cdot X \nonumber \\
&{}+ p7\cdot Z^3\cdot Y + p8\cdot Y^3\cdot Z + p9\cdot X^2\cdot Y^2 \nonumber \\
&{}+ p10\cdot X^2\cdot Z^2 + p11\cdot Z^2\cdot Y^2 + p12\cdot X^3 \nonumber \\
&{}+ p13\cdot Y^3 + p14\cdot Z^3 + p15\cdot X^2\cdot Y + p16\cdot Y^2\cdot X \nonumber \\
&{}+ p17\cdot X^2\cdot Z + p18\cdot Z^2\cdot X + p19\cdot Z^2\cdot Y \nonumber \\
&{}+ p20\cdot Y^2\cdot Z + p21\cdot X^2 + p22\cdot Y^2 + p23\cdot Z^2 \nonumber \\
&{}+ p24\cdot X\cdot Y + p25\cdot Y\cdot Z + p26\cdot X\cdot Z \nonumber \\
&{}+ p27\cdot X + p28\cdot Y + p29\cdot Z + p30,
\end{eqnarray}
where X, Y, Z represent the color $g-i$, [Fe/H], and log\,\textit{g}, respectively. 

\begin{figure*}[htbp]
\centering
\subfigure{
\includegraphics[width=0.49\textwidth]{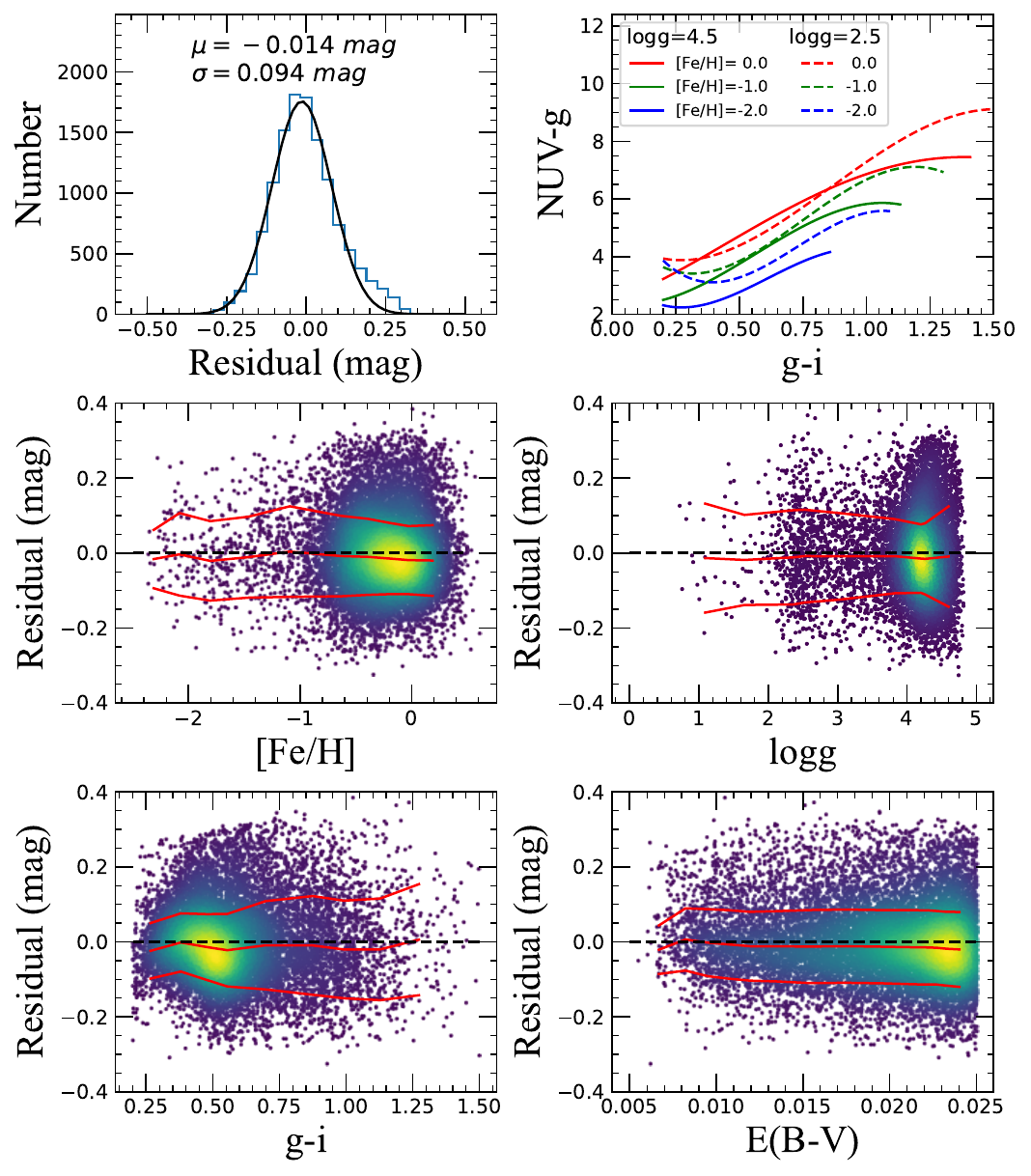}}
\subfigure{
\includegraphics[width=0.49\textwidth]{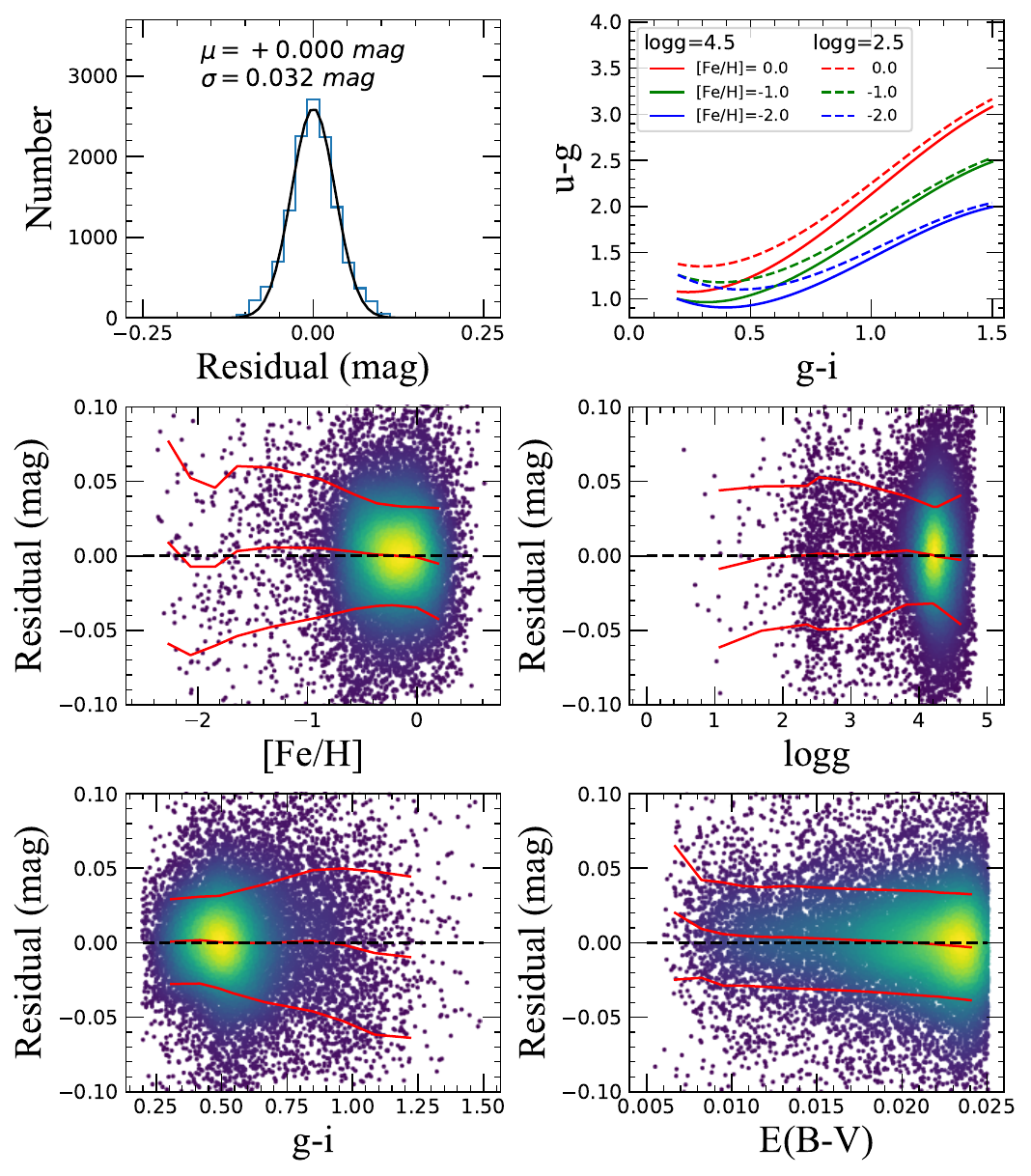}}
\caption{
Fitting residuals for the [Fe/H]- and log\,\textit{g}-dependent stellar loci based on observational data.
The left six panels are for the $NUV-g$ color, which contain the histogram distribution of fitting residuals, with the Gaussian fitting profile over-plotted in black, and the $NUV-g$ vs. $g-i$ stellar loci at different [Fe/H] values for dwarfs($\rm log\,\textit{g}=4.5$) and giants($\rm log\,\textit{g}=2.5$), and fitting residuals as a function of [Fe/H], log\,\textit{g}, $g-i$, and $E(B-V)$ with the median values and standard deviations over-plotted in red.
The right six panels are similar to the left ones but for the $u-g$ color. The points are color-coded by their number density.
}
\label{fig:obs_fit}
\end{figure*}

The fitting residuals for the $NUV-g$ and $u-g$ colors are shown in Figure\,\ref{fig:obs_fit}. The mean values of the fitting residuals are also close to 0.0 mag for all colors. The standard deviations are 0.094, 0.032, 0.004, 0.004\,mag for the colors $NUV-g$, $u-g$, $g-r$, and $i-z$, respectively. 
All  residuals exhibit no discernible trends with $g-i$, [Fe/H], log\,\textit{g}, and $E(B-V)$.

As in Section \ref{sec:model_data}, we estimate the photometric metallicities and log\,\textit{g} by the minimum $\chi^2$ technique. Adopting a 0.01\,dex step for [Fe/H] from $-$3.0 to $+$1.0 and log\,\textit{g} from 0.0 to 5.0, this produces a 401$\times$501 $\chi^2$ grid. Note that the parameter space bounded by $0.5<g-i<0.65$ and $0<\log\,\textit{g}<2.5$ is devoid of stars.

\begin{figure*}[htbp]
\centering
\includegraphics[width=13cm]{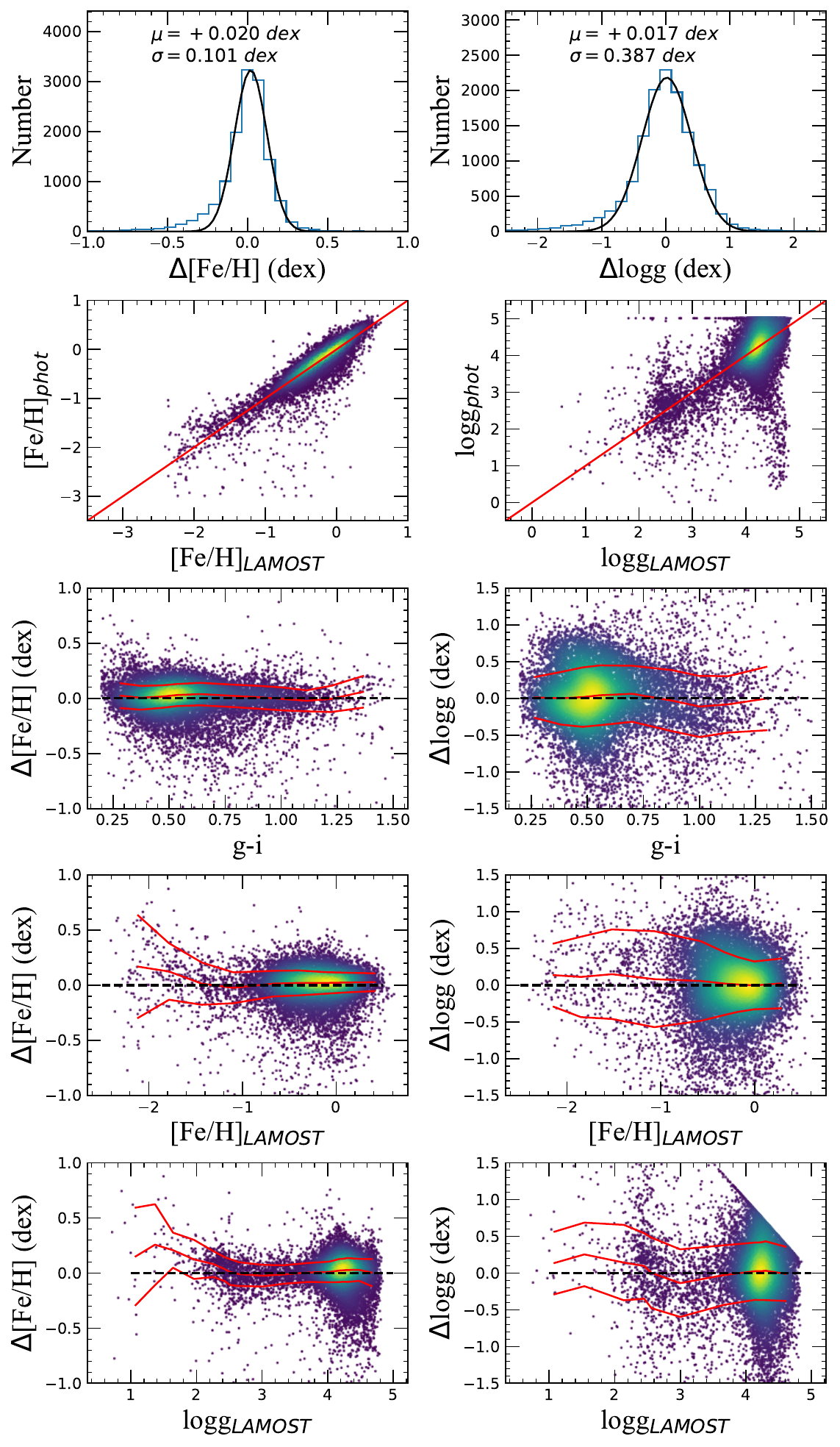}
\caption{Similar to Figure\,\ref{fig:simu_result} but for results based on observational data. 
}
\label{fig:obs_result}
\end{figure*}

\begin{figure*}[htbp]
\centering
\subfigure{
\includegraphics[width=0.45\textwidth]{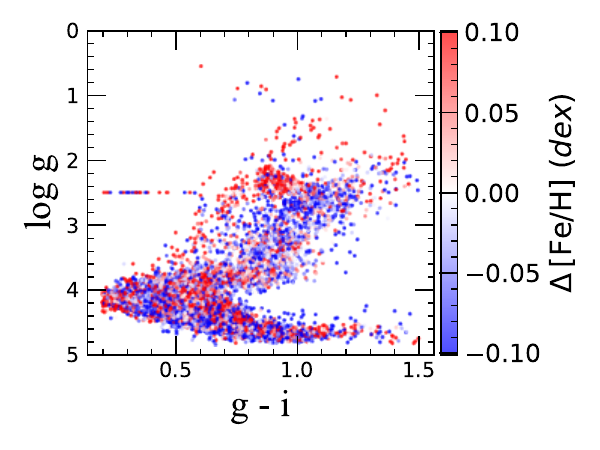}}
\subfigure{
\includegraphics[width=0.45\textwidth]{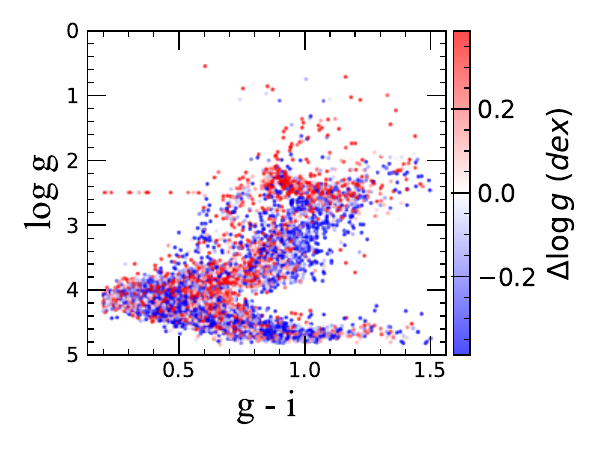}}
\caption{
The distribution of $\Delta[\rm Fe/H]$ and $\Delta \rm log\,\textit{g}$ in the H-R diagram. The color bars represent $\Delta[\rm Fe/H]$ and $\Delta \rm log\,\textit{g}$, respectively.
}
\label{fig:resi_HRD}
\end{figure*}

The results are shown in Figure\,\ref{fig:obs_result}. We have achieved a typical precision of 0.101\,dex for [Fe/H] and 0.387\,dex for log\,\textit{g}, respectively. Both residuals are flat as functions of $g-i$, $\rm [Fe/H]_{LAMOST}$, and $log\,\textit{g}_{LAMOST}$. 
A strong one-to-one correlation is found between $\rm [Fe/H]{phot}$ and $\rm [Fe/H]_{LAMOST}$, except for a few outliers far below the red line, which are likely due to stellar activity in the $NUV$-band, and a downward tail just below the red line corresponding to the left tail of the histogram, which is primarily caused by unrecognized binaries. 
It should be noted that the $\rm [Fe/H]_{phot}$ error (0.101 \, divisible) from the observational data is much smaller than that (0.183 \, divisible) from theoretical data with 0.01\,mag errors. 
For the $\rm log\,\textit{g}_{phot}$ vs. $\rm log\,\textit{g}_{LAMOST}$ panels, most sources exhibit good agreement, except for a small fraction of dwarfs at the bottom right that are incorrectly classified as giants. 
However, as shown in Figure\,\ref{fig:resi_HRD}, both $\Delta [\rm Fe/H]$ and $\Delta \rm log\,\textit{g}$ exhibit systematic trends in the H-R diagram. Yet, these deviations are small, remaining within $\pm 1\sigma$, and are mainly caused by red-clump stars.

We note that we find a positive correlation between $\Delta[\rm Fe/H]$ and $\Delta log\,\textit{g}$: an overestimate of 1 dex in log\,\textit{g} causes about an overestimate of 0.2 dex in [Fe/H]. The parameter residuals in our results cannot be attributed to systematics in the LAMOST parameters, as any possible systematic uncertainties are much smaller than the systematic trends identified in our analysis.
This inspires us to explore other methods for giant-dwarf classifications in order to achieve better photometric-metallicity estimates.

\section{Test of the giant-dwarf loci method  with observational data} \label{sec:select_giant}

In this Section, following \citeauthor{Zhang_2021} (\citeyear{Zhang_2021}, Paper IV), we use the giant-dwarf loci method to classify giants and dwarfs and estimate their photometric metallicities using CSST-like photometry. The stellar locus differences between dwarfs and giants make the classifications feasible. 
First, we divide the observational data into dwarfs and giants on the basis of an empirical boundary in the H-R diagram. The boundary is defined as:
\begin{eqnarray}\label{boundary}
\log g = -(g-i)^2 + 2.45\times(g-i) +2.72
\end{eqnarray}

Then, we obtain the [Fe/H]-dependent stellar loci for the dwarf and giant samples, respectively. 
For dwarfs, a 4th-order polynomial is adopted for the $NUV-g$ color, and 3rd-order polynomials are adopted for the colors $u-g$, $g-r$, and $i-z$.
For giants, a 4th-order polynomial is adopted for the $u-g$ color, and 3rd-order polynomials are adopted for the colors $NUV-g$, $g-r$, and $i-z$.
The 4th-order polynomials are expressed as:
\begin{eqnarray}\label{D2O4}
&Color = \nonumber\\
&p0\cdot X^4 + p1\cdot Y^4 + p2\cdot X^3\cdot Y + p3\cdot Y^3\cdot X \nonumber \\
&{}+ p4\cdot X^2\cdot Y^2 + p5\cdot X^3 + p6\cdot Y^3 + p7\cdot X^2\cdot Y \nonumber \\
&{}+ p8\cdot Y^2\cdot X + p9\cdot X^2 + p10\cdot Y^2 + p11\cdot X\cdot Y \nonumber \\
&{}+ p12\cdot X + p13\cdot Y  + p14,
\end{eqnarray}

\begin{figure*}[htbp]
\centering
\includegraphics[width=14cm]{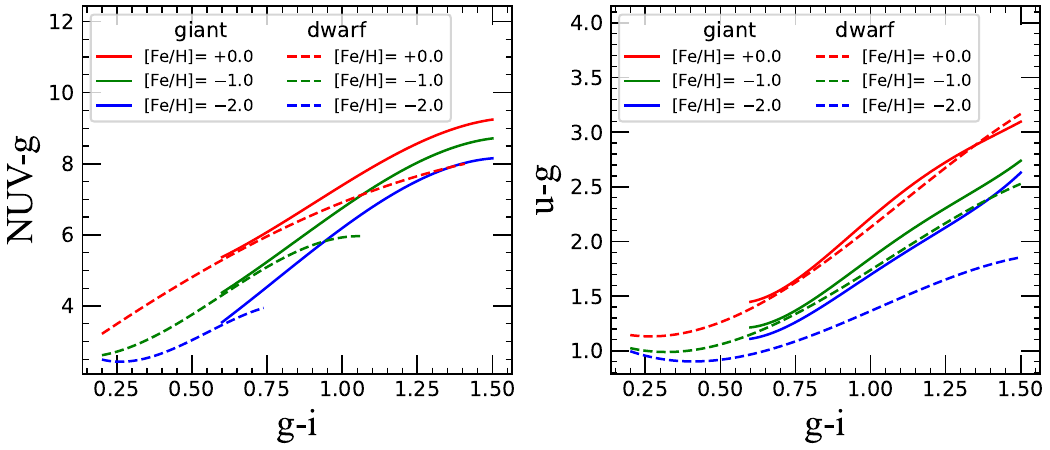}
\caption{Comparison of stellar loci for $NUV-g$ (left panel) and $u-g$ (right panel) vs. $g-i$ at different [Fe/H] values for dwarfs and giants. }
\label{fig:s_l_dif}
\end{figure*}

The 3rd-order polynomials are expressed as:
\begin{eqnarray}\label{D2O3}
&Color = \nonumber\\
& p0\cdot X^3 + p1\cdot Y^3 + p2\cdot X^2\cdot Y + p3\cdot Y^2\cdot X \nonumber \\
&{} + p4\cdot X^2 + p5\cdot Y^2 + p6\cdot X\cdot Y + p7\cdot X + p8\cdot Y  + p9, \nonumber \\
\end{eqnarray}
where X, Y represent the $g-i$ color and [Fe/H], respectively. 
Figure\,\ref{fig:s_l_dif} show the locus differences between dwarfs and giants for $NUV-g$ vs. $g-i$ and $u-g$ vs. 
$g-i$. At the same $g-i$ color, giants are generally bluer in $NUV-g$ and $u-g$; these differences become larger for redder and more metal-poor stars. 

Based on both sets of stellar loci, the minimum $\chi^2$ technique mentioned in Section \ref{subsec:feh_depen} is performed to estimate the photometric metallicities for the whole sample. The value of [Fe/H] varies from $-$3.0 to $+$1.0 with a step size of 0.01\,dex, resulting in 401 $\chi^2$ values for each set of stellar loci. 
Two estimates of photometric metallicity, $\rm [Fe/H]_{dwarf}$ and $\rm [Fe/H]_{giant}$, along with two minimum $\chi^2$ values, $\chi^2_{min, dwarf}$ and $\chi^2_{min, giant}$, are obtained for each source. 

\begin{figure*}[htbp]
\centering
\subfigure{
\includegraphics[width=7cm]{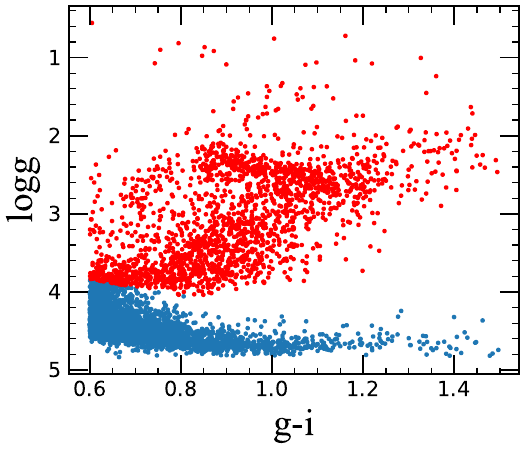}}
\subfigure{
\includegraphics[width=7cm]{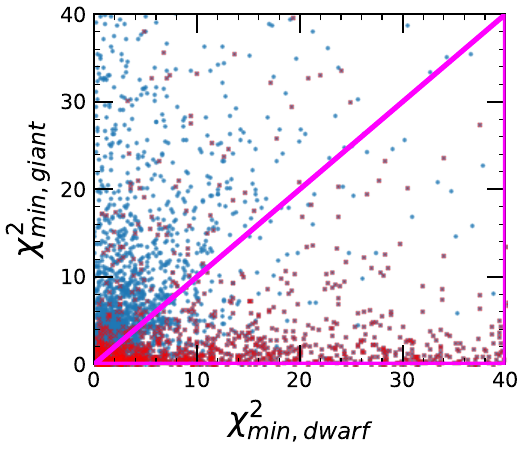}}
\caption{Left panel: The H-R diagram of the sample with $g-i>0.6$. The red and blue dots represent giants and dwarfs, respectively. The boundary of dwarfs (blue dots) and giants (red dots) is defined by Equation\,\ref{boundary}. Right panel: Values of $\chi^2_{min, dwarf}$ against $\chi^2_{min, giant}$. }
\label{fig:minchi2_tech}
\end{figure*}

\begin{figure*}[htbp]
\centering
\includegraphics[width=14cm]{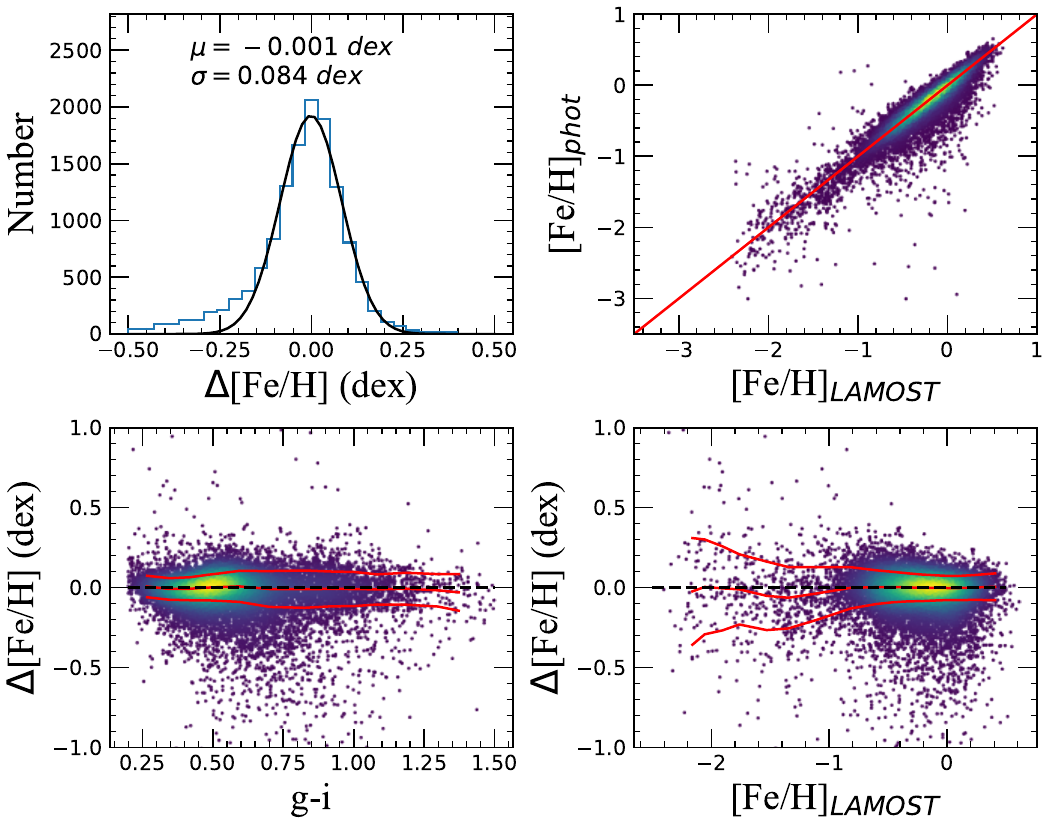}
\caption{Top left: Histogram distribution of the residuals $\Delta [\rm Fe/H]$ for the giant-dwarf loci method, with the Gaussian fitting profile over-plotted in black. Top right: comparison of $\rm [Fe/H]_{phot}$ with $\rm [Fe/H]_{LAMOST}$.
Bottom panels: The residuals $\rm \Delta [Fe/H]$, as a function of $g-i$ color and $\rm [Fe/H]_{LAMOST}$. The points are color-coded by their number density.
}
\label{fig:minchi2_feh}
\end{figure*}

Finally, we classify dwarfs and giants by comparing $\chi^2_{min, dwarf}$ and $\chi^2_{min, giant}$. Sources with $\chi^2_{min, giant} < \chi^2_{min, dwarf}$, which fall in the purple triangle region in the right panel of Figure\,\ref{fig:minchi2_tech}, are taken as giant candidates, and the values of $\rm [Fe/H]_{giant}$ are adopted; otherwise they are classified as dwarfs. Note that all stars with $g-i<0.6$ are selected as dwarfs.
The final results for $\rm [Fe/H]_{phot}$ are shown in Figure\,\ref{fig:minchi2_feh}. The typical precision of $\rm [Fe/H]_{phot}$ is 0.084\,dex.
A strong one-to-one correlation is also found between $\rm [Fe/H]_{phot}$ and $\rm [Fe/H]_{LAMOST}$, in addition to a few outliers far below the red line, possibly due to stellar activity in the $NUV$-band, and a downward tail just below the red line corresponding to the left tail of the histogram, which is likely caused by binaries.
The residuals are flat with respect to the color $g-i$ and $\rm [Fe/H]_{LAMOST}$.

\begin{figure}[htbp]
\centering
\includegraphics[width=\linewidth]{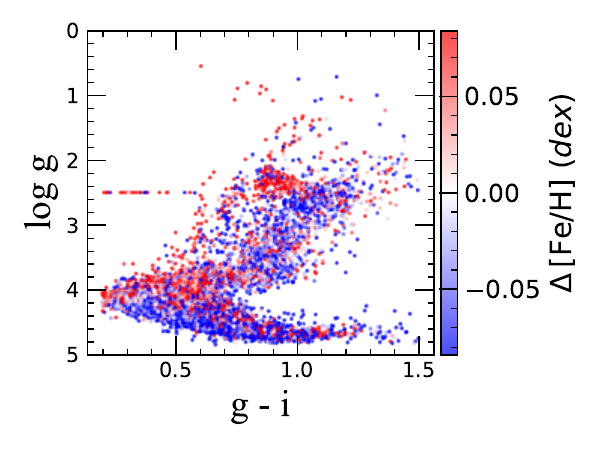}
\caption{Distribution of the residuals $\Delta[\rm Fe/H]$ in the H-R diagram. The color bar represents $\Delta[\rm Fe/H]$. 
}
\label{fig:resi_feh_HRD_method2}
\end{figure}

Compared to the results in Section \ref{sec:observ_data}, the $\rm [Fe/H]_{phot}$ estimates here are more precise. This occurs primarily because both giants and dwarfs exhibit strong correlations between color and log\,\textit{g}. 
As shown in Figure \ref{fig:resi_feh_HRD_method2}, the $\Delta [\rm Fe/H]$ residuals on the H-R diagram display two systematic biases — one for red-clump stars and the other for turn-off stars, but are confined within the $\pm 1\sigma$ range of 0.084\,dex. Hence, 
log\,\textit{g} effects should be considered, if available, for more precise photometric metallicities.

\begin{figure*}[htbp]
\centering
\includegraphics[width=16cm]{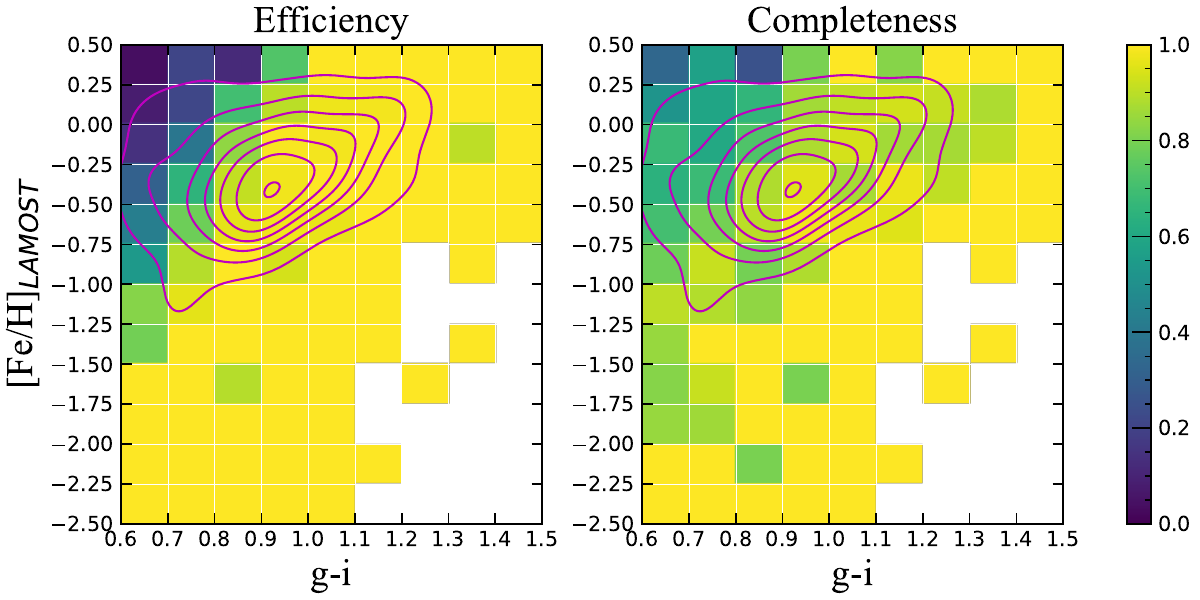}
\caption{Selection efficiency (left panel) and completeness (right panel) of giants as a function of $g-i$ color and $\rm [Fe/H]_{LAMOST}$ for the giant-dwarf loci method. The contour profile represented the number density of giants.}
\label{fig:minchi2_classi}
\end{figure*}

We now consider the sources with $g-i>0.6$ to test the dwarf and giant classifications. 
The dwarfs and giants in our test sample and their classifications are shown in Figure\,\ref{fig:minchi2_classi}. From inspection, most giants are classified correctly. Statistically, the giant-dwarf loci method performs better at distinguishing giants from dwarfs. 

To further evaluate our classifications, we divide the sample into two-dimensional bins of $g-i$ and $\rm [Fe/H]_{LAMOST}$. The ranges are 0.6 to 1.5 in color and $-$2.5 to 0.5 in metallicity, with steps of 0.1\,mag and 0.25\,dex, respectively. For each bin, the selection completeness, defined as the ratio of the number of giants selected as candidates to the number of all giants in the bin, and the selection efficiency, defined as the ratio of the number of giants selected to the number of all giant candidates in the bin, are calculated. The two-dimensional distributions of the completeness and efficiency of the selection in the $g-i$ versus $\rm [Fe/H]_{LAMOST}$ panels are shown in Figure\,\ref{fig:minchi2_classi}, with contours indicating the number density of giants. Bins with no stars are colored white. From inspection, both efficiency and completeness exhibit increasing trends as [Fe/H] decreases and $g-i$ color increases. The efficiency is nearly 100\% for metal-poor giants with $[\rm Fe/H]<-1.0$ or redder giants with $g-i>1.0$, decreasing towards more metal-rich and bluer giants. However, note that there are very few metal-rich giants with $g-i<0.8$, as shown in the contour. The completeness shows smaller variations, with a typical value of 80\%. We conclude that this method can select giants, especially metal-poor or redder giants, with extremely high efficiency and completeness.

\section{Summary} \label{sec:summary}
In this work, we present two methods to estimate stellar parameters from CSST-like photometry, using theoretical data synthesized from the BOSZ spectra as well as observational data constructed from GALEX GR5 MIS $NUV$-band photometry, synthetic SDSS $ugriz$ magnitudes from ``corrected” Gaia XP Spectra, and spectroscopic parameters from LAMOST DR8.

The first method uses the [Fe/H]- and log\,\textit{g}- dependent stellar loci of colors $NUV-g$, $u-g$, $g-r$, $i-z$, and $z-y$ to simultaneously estimate metallicity and log\,\textit{g}.
Our test with theoretical data (without photometric errors) results in a precision of 0.088\,dex and 0.083\,dex for [M/H] and log,\textit{g}, respectively.
With the assumed 0.01\,mag photometric errors, the precision decreased to 0.183\,dex, and 0.176\,dex for [M/H] and log\,\textit{g}, respectively, primarily due to the degeneracy in metallicity and log\,\textit{g}.
Our test with the observational data, albeit with larger photometric errors, results in a precision of 0.101\,dex and 0.387\,dex for [Fe/H] and log\,\textit{g}, respectively, due to the strong correlation between stellar colors and log\,\textit{g} in real data.

The second approach is the giant-dwarf loci method, which is based on the different metallicity-dependent stellar loci between dwarfs and giants to obtain classifications and metallicity estimates. With the same observational sample, the second method results a slightly better [Fe/H] precision of 0.084\,dex compared to the first method, due to the stronger constraints imposed on log\,\textit{g}. The giant-dwarf loci method also performs well in distinguishing giants from dwarfs, particularly for metal-poor or redder giants. Both the selection efficiency and completeness is close to 100\% for metal-poor giants with $[\rm Fe/H]<-1.0$ or redder giants with $g-i>1.0$.

We conclude that the CSST-like $NUV$- and $u$-bands are highly sensitive to both metallicity and log,\textit{g}, allowing us to estimate these two parameters simultaneously and precisely when both bands are used. 
To achieve high precision, the degeneracy between metallicity and log\,\textit{g} has to be carefully taken into account. The tests presented in this work demonstrate the potential of the CSST and the power of stellar-loci based methods in achieving such goals, paving the way for stellar parameter estimates for many billions of stars with future CSST data.

\clearpage

\begin{acknowledgments}

The authors thank the referee for his/her suggestions that improved the clarity of our presentation.
This work is supported by the National Natural Science Foundation of China through the project NSFC 12222301, 12173007,
the National Key Basic R\&D Program of China via 2024YFA1611901 and
2024YFA1611601.  
This work is also supported by the China Manned Space Program
with grant no. CMS-CSST-2025-A13.
T.C.B. acknowledges partial support from grants PHY 14-30152; Physics Frontier Center/JINA Center for the Evolution of the Elements (JINA-CEE), and OISE-1927130; The International Research Network for Nuclear Astrophysics (IReNA), awarded by the US National Science Foundation, and DE-SC0023128; the Center for Nuclear Astrophysics Across Messengers (CeNAM), awarded by the U.S. Department of Energy, Office of Science, Office of Nuclear Physics. 
His participation in this work was initiated
by conversations that took place during a visit to China
in 2019, supported by a PIFI Distinguished Scientist award
from the Chinese Academy of Science. 

This work has used data from the European Space Agency (ESA) Gaia mission (\url{https://www.cosmos.esa.int/gaia}), processed by the Gaia Data Processing and Analysis Consortium (DPAC, \url{https:// www.cosmos.esa.int/web/gaia/dpac/ consortium}). Funding for the DPAC has been provided by national institutions, in particular, the institutions participating in the Gaia Multilateral Agreement. 
Guoshoujing Telescope (the Large Sky Area Multi-Object Fiber Spectroscopic Telescope LAMOST) is a National Major Scientific Project built by the Chinese Academy of Sciences. Funding for the project has been provided by the National Development and Reform Commission. LAMOST is operated and managed by the National Astronomical Observatories, Chinese Academy of Sciences.
We acknowledge the invaluable contribution of GALEX (NASA's Galaxy Evolution Explorer) in providing the dataset used in this research.

\end{acknowledgments}

\bibliographystyle{aasjournal}
\bibliography{main}

\end{document}